\documentclass[12pt,preprint]{aastex}

\slugcomment{}

\shorttitle{The Edge of the Milky Way Stellar Disk Revealed Using Clump Giant Stars as Distance Indicators}
\shortauthors{Minniti et al.}

\begin{document}

\title{The Edge of the Milky Way Stellar Disk Revealed Using Clump
  Giant Stars as Distance Indicators}

\author{D.~Minniti\altaffilmark{1,2,3}, R.~K.~Saito\altaffilmark{1},
  J.~Alonso-Garc\'{i}a\altaffilmark{1}, P.~W.~Lucas\altaffilmark{4}
  and M.~Hempel\altaffilmark{1}}
\affil{$^1$Departamento de Astronom\'{\i}a y Astrof\'{\i}sica,
  Pontificia Universidad Cat\'{o}lica de Chile, Vicu\~na Mackenna
  4860, Casilla 306, Santiago 22, Chile \\
$^2$Vatican Observatory, Vatican City State V-00120, Italy\\
$^3$European Southern Observatory, Vitacura 3107, Santiago, Chile\\
$^4$Centre for Astrophysics Research, University of Hertfordshire, College Lane,
Hatfield AL10 9AB, UK
}

\begin{abstract}
We use the clump giants of the disk as standard candles calibrated
from Hipparcos parallaxes in order to map their distribution with two
new near-IR surveys of the Galactic plane: UKIDSS-GPS and VVV.  We
explore different selection cuts of clump giants.  We conclude that
there is an edge of the stellar disk of the Milky Way at $R=13.9 \pm
0.5$ kpc along various lines of sight across the galaxy. The effect of
the warp is considered, taking fields at different longitudes and
above and below the plane.  We demonstrate that the edge of the
stellar disk of the Milky Way can now be mapped in the near infrared
in order to test different models, and to establish our own place
within the galaxy.
\end{abstract}

\keywords{Galaxy: disk --- Galaxy: structure --- Stars: distances ---
  Stars: late-type}

\section{Introduction}

Most spiral galaxies have rather sharp edges in their stellar disks
\citep{1979A&AS...38...15V, 2007A&A...466..883V}. The situation for
the Milky Way is unclear. The disk of our galaxy is well represented
by an exponential density profile \citep{1970ApJ...160..811F}, with
some evidence for a stellar disk cut-off at a radius of $\sim14$~kpc
towards the anticentre region \citep{1992ApJ...400L..25R}, but other
claims for a warped and flaring stellar disk out to 23~kpc
\citep{2006A&A...451..515M}. Based upon the scaling relation
established by other galaxies, the radial cut off should be between 10
and 25~kpc \citep{2004ASSL..319..713P, 2004MNRAS.355..143K}. Here we
report observations of ``clump red giant'' stars, which are good
distance indicators \citep{2000ApJ...539..732A, 1998ApJ...494L.219P},
along 10 lines of sight in the disk of the Milky Way. The data reveal
an edge to the disk at $R=13.9\pm0.6$~kpc. Why there should be a sharp
edge to the stars, while the gas profile is much more extended and
does not show such an edge \citep{2009ARA&A..47...27K}, remains
unclear, but a critical test would be the measurement of the gas and
star density in situ \citep{1989ApJ...344..685K}.

There are two new deep surveys of the Milky Way disk in the
near-infrared: the UKIDSS Galactic Plane Survey (GPS), that is mapping
the Northern disk \citep{2008MNRAS.391..136L}, and the VISTA Variables
in the {\it V\'{i}a L\'actea} (VVV) Survey, that is mapping 520 square
degrees in the Southern disk and bulge of our Galaxy
\citep{2010NewA...15..433M}. While previous observations in the
Galactic plane have been limited by crowding, source dimness, and
interstellar extinction, these new public surveys allow us to pierce
through the whole disk, reaching red giants in the horizontal branch
(clump giants) located at the other side of our galaxy for the first
time in different near-IR pass-bands. The VVV survey observes in the
$Z$ (0.87 microns), $Y$ (1.02 microns), $J$ (1.25 microns), $H$ (1.64
microns), and $K_{\rm s}$ (2.14 microns) bands, while the UKIDSS-GPS
is observing in the $J$ (1.25 microns), $H$ (1.64 microns), and $K$
(2.14 microns) bands; with the photometry uniformly calibrated in the
2MASS system \citep{2006AJ....131.1163S}.

  \begin{figure*}[ht]
  \includegraphics[bb=.5cm .7cm 5cm 5cm,scale=1.8]{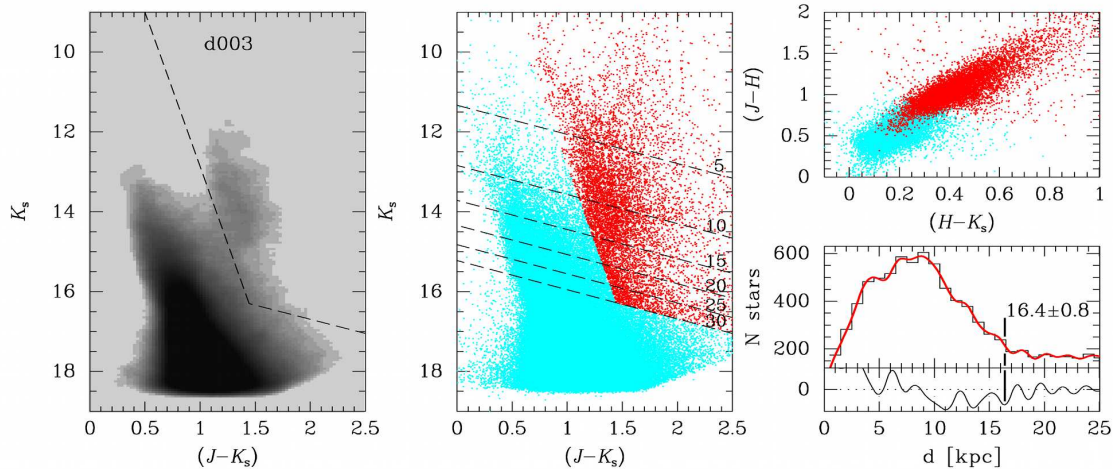}
   \caption{Selection of giant stars used in the distance
     calculation. The left-hand panel shows the colour-magnitude
     diagram (CMD) of VVV field d003 as a density map. The adopted
     color-cut is shown, which is similar to the cut for other fields,
     even though the location of the edge feature is relatively
     insensitive to these cuts. The CMD in the central panel and the
     colour-colour diagram in the top-right panel show in dark gray
     the selection of giant stars used in the distance
     calculation. Dashed lines in the CMD mark the distance in kpc for
     the selected stars according with equation 2. The bottom-right
     panel shows the distance distribution and respective derivative
     curve, with the outermost the minimum derivative marked.}
  \label{cmd}
  \end{figure*}

Clump giants are excellent tools to study the structure of our galaxy
because of two main reasons, discussed below: (i) they are good
distance indicators, and (ii) they are good tracers of the mass in the
disks of galaxies. The UKIDSS-GPS and VVV Surveys give the largest
homogeneous census of clump giants out to well beyond the extent of
the Milky Way stellar disk. The limiting DoPhot
\citep{1993PASP..105.1342S} magnitude of single epoch VVV images
processed by the Cambridge Astronomical Survey Unit (CASU) pipeline
v1.0 is $K_{\rm s}=17.5$ in the disk fields, and the DoPhot photometry
on the combined epochs allow to reach approximately $J=20$~mag, and
$K_{\rm s}=18.5$~mag (Fig.~\ref{cmd}), slightly fainter than the UKIDSS-GPS,
and much fainter than the near-IR surveys 2MASS and DENIS, for which
the limit is $K_{\rm s}=14.3$~mag \citep{2006AJ....131.1163S}. The
distance probed along the line of sight depends on the reddening of
the fields. For example, in disk fields with low extinction ($A_{\rm
  V}<3$~mag) the UKIDSS-GPS and VVV surveys would see clump giants beyond
50~kpc. Therefore we can search for the edge of the disk of our galaxy
using well calibrated standard candles.

\section{Observational Data}

The clump giants with known parallaxes measured by the Hipparcos
satellite are the best collectively calibrated standard candles
\citep{1998ApJ...494L.219P}. 

Figure~\ref{cmd} shows a colour-colour and colour-diagram magnitude
for the VVV field d003. In order to select the giant stars a colour
cut was applied in the $K_{\rm s}$ $vs.$ ($J-K_{\rm s}$), as shown in left-hand
and central panels of Fig.~\ref{cmd}. Not all selected stars are clump
giants, but the presence of sub-giant branch and red-giant branch
stars, or even main-sequence stars at fainter magnitudes, contribute
to the smooth underlying background. These different population do not
affect the location of any sharp edges of the stellar distribution,
which can however be well defined by the clump giants. The effect of
reddening in the selection of clump giants was also tested using
different selection criteria both in the colour-magnitude and
colour-colour diagrams, finding consistent results.

An accurate $K$-band calibration of the
red clump giant luminosity was obtained and applied to the red giant
clump of the Galactic bulge \citep{2000ApJ...539..732A}, and of the
Large Magellanic Cloud \citep{2002ApJ...573L..51A}. Our magnitudes are
in the 2MASS magnitude system, and we transformed the $K$-band
magnitudes of clump giants of Hipparcos \citep{2000ApJ...539..732A} to
$K_{\rm s}$ magnitudes using $K= K_{\rm s}+0.044$. The resulting zero
point differences should be less than 0.02 magnitudes
\citep{2002ApJ...573L..51A}.  For the Ks photometry, the distance
modulus to a red clump giant in the outer disk would be:
\begin{equation}
\mu = K_{\rm s}- \frac{A_{K_{\rm s}}}{(A_J-A_{K_{\rm s}})}~\left[(J-K_{\rm
    s})-(J-K_{\rm s})_0\right]-M_{K_{\rm s}}
\end{equation}where we adopted $A_{K_{\rm s}}/(A_J-A_{K_{\rm s}})=0.73$, 
$(J-K_{\rm s})_0=0.70\pm 0.05$ and $M_{K_{\rm s}}=-1.65\pm 0.03$ as
the mean values for the red clump giants of the Milky Way disk
\citep{2002ApJ...573L..51A}. There should be negligible metallicity
dependence of these mean values because we are looking at a stellar
population of the Milky Way disk that should be similar to that of the
Solar neighbourhood where Hipparcos distances of clump giants were
calibrated.  Adopting these means magnitudes and colours, and the
reddening coefficients \citep{1989ApJ...345..245C} yields:
\begin{equation}
\mu = -5+5\,{\rm log}\,d({\rm pc}) = K_{\rm s}-0.73\,(J-K_{\rm s})+2.16\,.
\end{equation}

  \begin{figure}[ht]
  \includegraphics[bb=-6cm 4cm 25cm 21.3cm,scale=0.47]{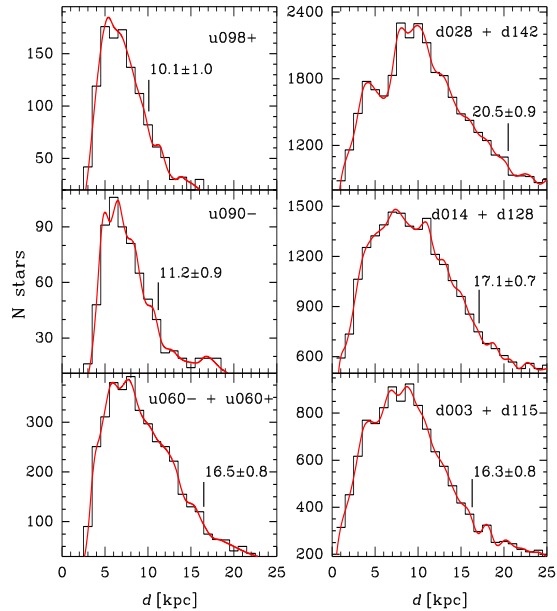}
   \caption{Distance distribution for clump giant stars along
     different lines of sight.  The distance measured to the edge of
     the Milky Way stellar disk is labelled for each field. The thin
     solid line is the smoothed distribution. These stellar
     distributions are rich with information about the disk
     structure. For example, we discarded the presence of distant star
     clusters, where the red giant clump could mask an edge feature,
     but see some over-density fluctuations that may be due to spiral
     arms.}
  \label{dist}
  \end{figure}

Using this equation we computed the distance modulus (and distance in
kpc) for every single clump giant candidate in the fields. For the UKIDSS-GPS
$K$-band data the computations are the same, but with
$A_K/(A_J-A_K)=0.68$, $(J-K)_0=0.66\pm 0.04$ and $M_K=-1.61\pm 0.03$.

In order to determine the disk edge we computed the distribution in
distance for the clump giants in each field. Thus we coadded the
distance distribution of fields located above and below the plane at
same Galactic longitude. The coadded distance distribution for each
longitude was analyzed non-parametrically using the local likelihood
density estimation method \citep{1996_Loader}. Finally, the derivative
method was applied for the distribution curve. This method is simple
and robust, and can be used to detect the structures in the distance
distribution. We defined as the disk edge the outermost point of
minimum derivative in each case. The uncertainties were calculated
with a Monte Carlo procedure, assuming Poisson errors.

\begin{figure*}[ht]
\includegraphics[bb=1cm -3.5cm 19cm 17cm,angle=-90,scale=0.58]{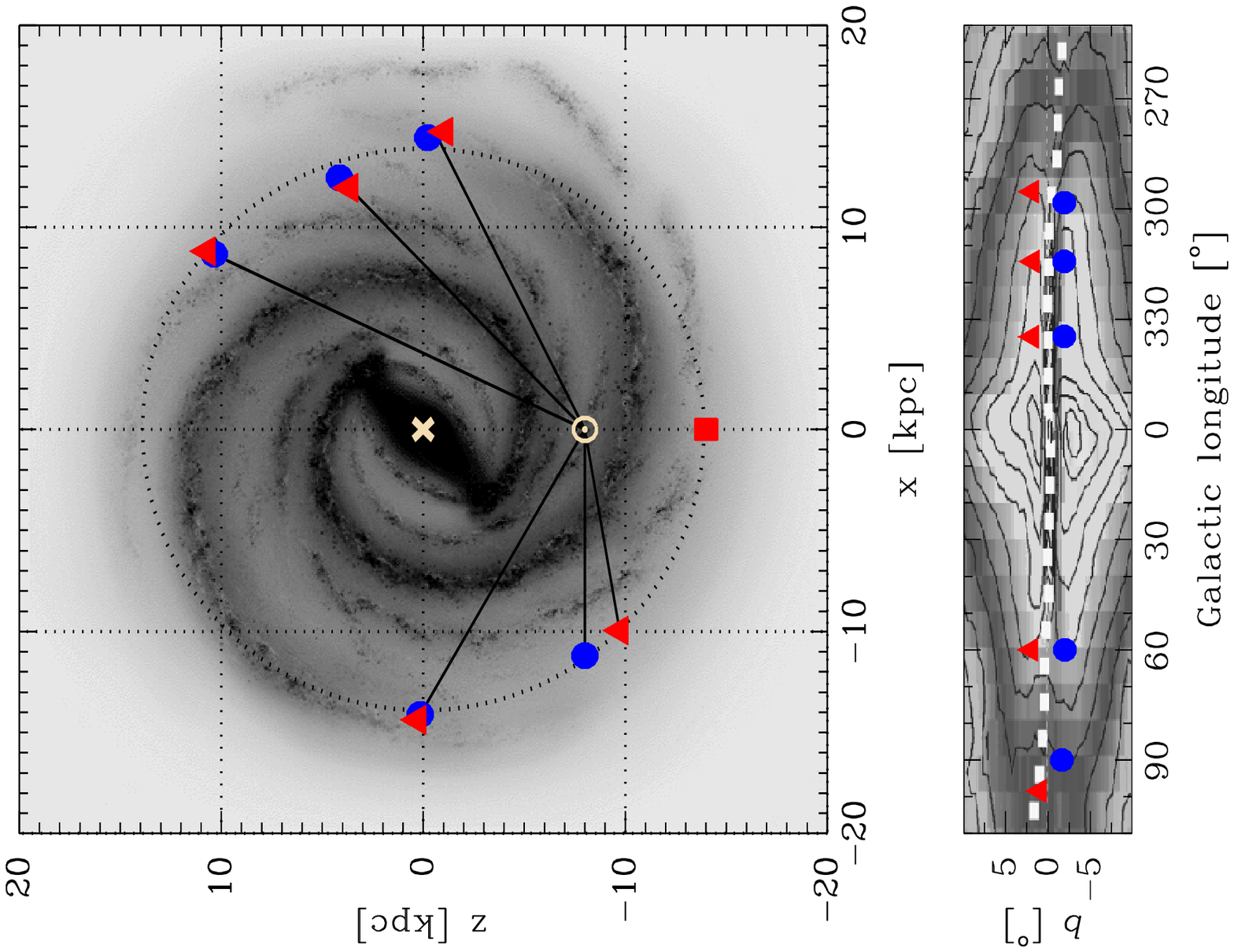}
 \caption{Measured positions of the edge of the Milky Way stellar
   disk. Top panel: Schematic face-on view of the Milky Way disk based
   on radio, infrared, and visible data \citep{2009PASP..121..213C},
   showing the positions of the edge of the stellar disk measured in
   different directions. The fields to the left are from the UKIDSS-GPS, the
   fields to the right are from the VVV survey, and the anticentre
   point at $R=14$~kpc is from \citep{2003A&A...409..523R,
     2009A&A...495..819R}. We adopted a distance to the galactic
   centre of $R_o=8$~kpc. This figure shows that a disk radius of
   $13.9\pm0.6$~kpc (dotted circle) fits the data very well. Bottom
   panel: Edge-on view of the Milky Way in Galactic $l$, $b$
   coordinates with the position of the warped stellar disk marked
   with white squares \citep{2006A&A...451..515M}. The positions of
   our fields are marked with triangles and circles, and the vertical
   scale has been stretched for clarity.}
\label{map}
\end{figure*}

The uncertainties due to the photometric errors do not produce any
trend in the calculations and are much smaller than the statistical
error. The resulting distance distributions along different lines of
sight are shown in Fig.~\ref{dist}. In all fields there is a density
drop in the clump giant distribution at a corresponding Galactocentric
distance of about 13.9~kpc (assuming $R_o=8$~kpc), as listed in
Table~\ref{table}.

We first noticed the sharp termination of the clump giant distribution
in the color-magnitude diagram of the VVV field d001 ($l, b =
295.4^{\circ}, -1.7^{\circ}$), a very heavily reddened area located in
the outskirts of the Carina star forming region
\citep{2010Msngr.141...24S}. Similar behavior is also found in
UKIDSS-GPS data for different Galactic longitudes
\citep{2008MNRAS.391..136L}. 

We studied low-extinction fields across the Milky Way disk (see
Table~\ref{table}). When these were available, we selected pairs of
fields located above and below the plane at the same Galactic
longitude, in order to account for the Galactic warp, because if the
disk is warped, the line of sight can leave the stellar disk before
this ends. For example, at $l=300^{\circ}$ the mean warp location is
$1^{\circ}$ below the plane \citep{1992ApJ...400L..25R}, so the
selected VVV field d003 is conveniently located to probe the full
extent of the disk. At most other longitudes explored here the warp is
absent, with the mean plane being at $b=0^{\circ}$ (Fig.~\ref{map}). We
have also explored a few other UKIDSS-GPS fields located at inner
longitudes that yield lower distances. These fields with
$l<60^{\circ}$ were discarded because they are heavily reddened and
crowded, and the photometry is not as deep as the rest, presumably due
to enhanced contamination from the near-side of the inner
bar. Systematic errors such as variations of the reddening law have
not been included.

\begin{figure}[ht]
\includegraphics[bb=-10cm 8cm 14cm 21.5cm,scale=0.43]{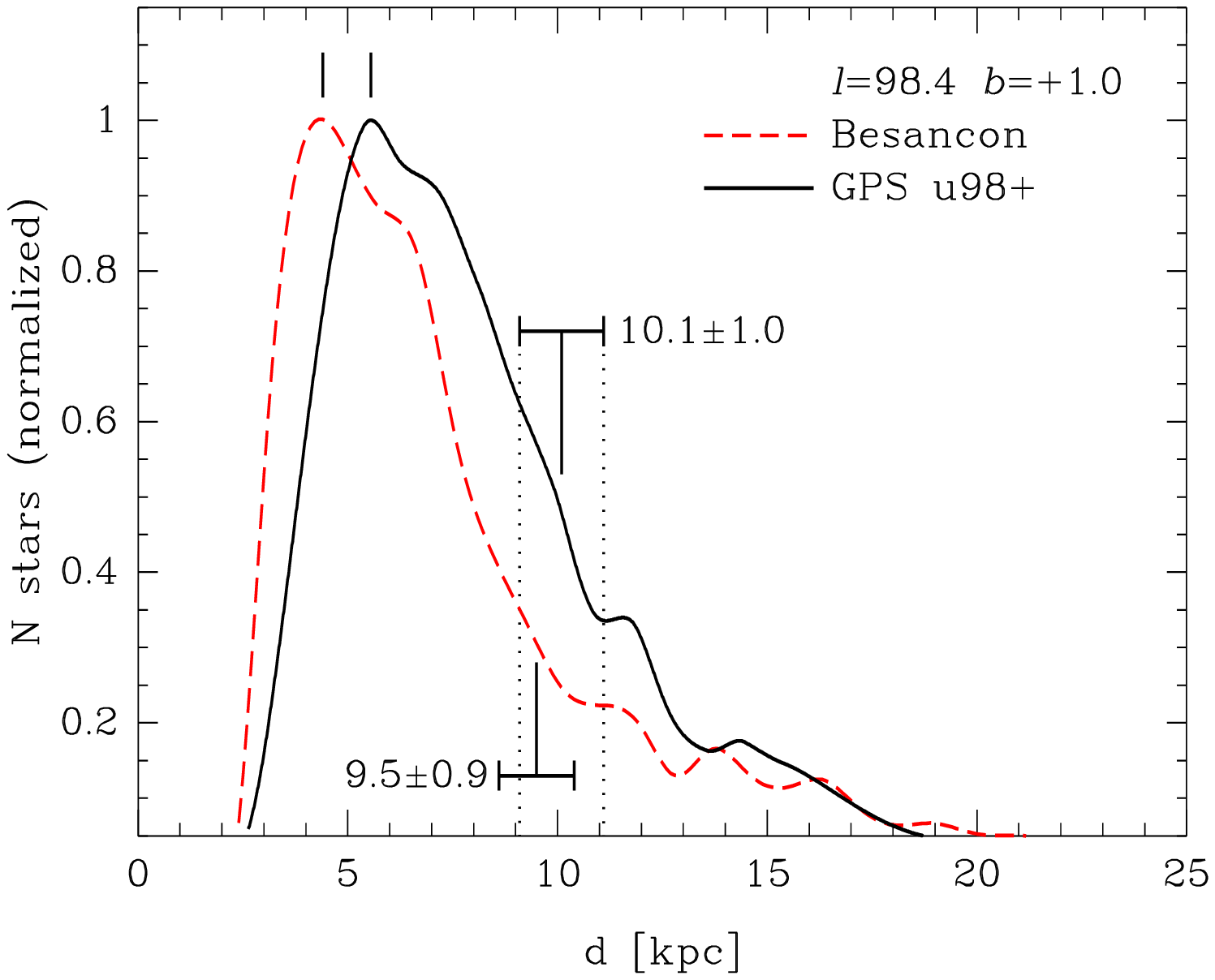}
 \caption{Comparison between data and model. Black solid line:
   Distance distribution for the field u098+ using UKIDSS-GPS
   data. Dashed red line: Simulation of the same field u098+ using the
   Besan\c{c}on Galactic model with a scale length of 2.6~kpc and an
   edge at 14~kpc. The model fits all features seen in the data
   distribution and the disk edge is detected at the expected place
   within the errors (see text). The distributions are arbitrarily
   normalized, and at shorter distances the discrepancy is due to
   photometric saturation.}
\label{model}
\end{figure}

It is important to note that the present results are independent of
models. They rely on basic assumptions such as that the Hipparcos
stellar sample is representative of the entire the Milky Way disk, and
that clump giants are reliable distance indicators and tracers of the
old and intermediate age populations. However, the Besan\c{c}on
Galactic model \citep{2003A&A...409..523R} with a scale-length of
2.4~kpc, including a flaring and warped outer disk, with a disk cutoff
at $R=14$~kpc seems to reproduce well the observed distance distributions (see
Fig.\ref{model}). Data and model show the same features, with a sharp
increase, the peak around 5~kpc and the exponential decrease. The
edge is detected in the data and model at the expected place within
the errors [$d=10.1$~kpc (corresponding to $R=13.9$~kpc), and
  $d=9.5$~kpc ($R=13.4$~kpc), respectively].

\section{Discussion and Conclusions}

A few studies have discussed the edge of the Milky Way towards the
anticentre region previously. The old Galactic disk apparently does
not extend beyond 14~kpc on the basis of optical $V$ vs. $(B-V)$
colour-magnitude diagrams and star counts \citep{1992ApJ...400L..25R},
which does not contradict the presence of other more distant young
stars. Models of the COBE-DIRBE infrared emission maps yield that the
edge of the disk is at 4~kpc from the Sun \citep[assuming
  $R_o=8.5$~kpc,][]{1996ApJ...468..663F}. DENIS star counts reveal the
cut off of the stellar disk at a Galactocentric distance of
$15\pm2$~kpc \citep{1996A&A...313L..21R}.

In contrast, star counts from 2MASS reveal no radial disk truncation
at 14~kpc \citep{2006A&A...451..515M}. Again, the depth of 2MASS is
not enough to reach large distances, which is possible with the VVV
and UKIDSS-GPS survey that reach 3-4 magnitudes fainter. More recently, it
was found that the 2MASS star counts are best fit if the external disk
is truncated at 12-14~kpc \citep{2009A&A...495..819R}, while early
A-type stars in the anticentre from the IPHAS survey reveal an
exponential disk out to 13~kpc, with a steeper decline beyond that
distance \citep{2010MNRAS.402..713S}. Most of the previous evidence
for cutoffs in the stellar distribution has been acquired at the
anticentre fields. The present deep exploration of different fields in
three galactic quadrants finds consistent results.

Even though we interpret this termination of the clump giant distribution 
as a truncation, we note that a break in the slope of the distribution 
would also be consistent with the data. It is difficult to measure
the sharpness of the cutoff, because the number of stars drop rapidly 
with distance, but as the surveys progress, we will have more fields 
in order to explore this issue.

\begin{table*}[ht]
\begin{center}
\caption{Selected GPS and VVV fields with relatively low extinction
  \citep{1993PASP..105.1342S} in separate directions across the Milky
  Way plane.} 
\vspace{12pt}
\begin{tabular}{l c c c c}
Field ID & $l$\,[$^{\circ}$]  & $b$\,[$^{\circ}$] & $d$\,[kpc] & $R$\,[kpc]\\
\hline
GPS-u098$+$ & $ 98.4$ & $+1.0$ & $10.1\pm1.0$ & $13.9\pm1.3$ \\
GPS-u090$-$ & $ 90.0$ & $-1.7$ & $11.2\pm0.9$ & $13.8\pm1.1$ \\
GPS-u060$-$ & $ 60.0$ & $-2.0$ & $16.3\pm0.7$ & $14.1\pm0.6$ \\
GPS-u060$+$ & $ 60.0$ & $+2.0$ & $16.6\pm0.8$ & $14.3\pm0.7$ \\
VVV-d028$ $ & $334.7$ & $-1.9$ & $20.3\pm0.7$ & $13.5\pm0.5$ \\
VVV-d142$ $ & $334.8$ & $+1.9$ & $20.7\pm1.0$ & $13.8\pm0.7$ \\
VVV-d014$ $ & $314.3$ & $-1.9$ & $17.4\pm0.9$ & $13.1\pm0.6$ \\
VVV-d128$ $ & $314.3$ & $+1.8$ & $16.7\pm1.3$ & $12.6\pm1.0$ \\
VVV-d003$ $ & $298.3$ & $-1.9$ & $16.4\pm0.8$ & $14.6\pm0.8$ \\
VVV-d115$ $ & $295.3$ & $+1.9$ & $16.3\pm0.8$ & $14.5\pm0.7$ \\
\hline
\label{table}
\end{tabular}
\end{center}
\end{table*}

This does not mean that one could not find other young stellar sources
beyond the distance that we measure. There are stars detected beyond
the edge of the old stellar disk. For example, there is a population
of distant young stars at $R=20$~kpc, in two fields located in the
third quadrant, slightly above the plane (at $b=7^{\circ}$ and
$b=4^{\circ}$), based on $UBV$ photometry
\citep{2010ApJ...718..683C}. We see no evidence of a large population
of clump giants at that distance in our fields.

$K$-band light traces mass in the disks of spiral galaxies
\citep{1993ApJ...418..123R}, and the integrated $K$-band luminosity is
dominated by red giants. Thus the clump giants are ideal tracers to
define overall features and structural parameters of our Milky Way,
like the edge of its disk. The red clump giants trace the old and
intermediate age populations, which are more uniformly distributed
than young stars, following the mass of the galactic disk.  However,
thick disk and halo giants are expected to be located beyond the edge
of the stellar thin disk, as have been detected by the SEGUE project
\citep{2010ApJ...714..663D}, contributing to the background that we
see in all fields beyond a Galactocentric distance of 13.9~kpc.
A similar result has been recently obtained independently by the
GLIMPSE team (Benjamin et al. 2011, in preparation).

Finally, we are able to determine not only the extension but also the
shape of the stellar disk of our galaxy for the first time. Asymmetric
features such as lopsidedness are common in spiral galaxies
\citep{1995ApJ...447...82R}. However, the results shown in
Fig.~\ref{map} suggest that the stellar disk of the Milky Way is not
significantly lopsided.

\acknowledgments

We acknowledge support by the FONDAP Center for Astrophysics, BASAL
CATA Center for Astrophysics and Associated Technologies, MILENIO
Milky Way Millennium Nucleus from MIDEPLAN, FONDECYT from CONICYT, and
the European Southern Observatory. RS acknowledges financial support
from CONICYT through GEMINI Project Nr. 32080016. We gratefully
acknowledge use of data from the VISTA telescope, and data products
from the Cambridge Astronomical Survey Unit, and the Two Micron All
Sky Survey (2MASS), which is a joint project of the University of
Massachusetts and IPAC/CALTECH, funded by NASA and NSF.

\end{document}